\documentclass[a4paper,10pt]{article}
\usepackage[utf8]{inputenc}

\usepackage{amssymb}
\usepackage{graphicx}
\usepackage{bm}
\usepackage{amsmath}
\usepackage{longtable}
\usepackage{lscape} 
\usepackage{color}
\usepackage[labelfont=bf,labelsep=period,justification=raggedright]{caption}
\usepackage{hyperref}

\title{{\bf Looplessness in networks is linked to trophic coherence}}
\author{Samuel Johnson$^{1\ast}$ and Nick S. Jones$^{2}$
\\
\small{$^{1}$Warwick Mathematics Institute, and Centre for Complexity Science,}\\
\small{University of Warwick, Coventry CV4 7AL, United Kingdom.}\\
\small{$^{2}$Department of Mathematics, Imperial College London,}\\
\small{London SW7 2AZ, United Kingdom.}\\
\small{$^\ast$E-mail:  S.Johnson.2@warwick.ac.uk}\\
}
\date{}

\begin{document}

\maketitle

\begin{abstract}
Many natural, complex systems are remarkably stable thanks to an absence of feedback acting on their elements.
When described as networks, these exhibit few or no cycles, and associated matrices have small leading eigenvalues.
It has been suggested that this architecture can confer advantages to the system as a whole, 
such as `qualitative stability', 
but this observation does not in itself explain how a loopless structure might arise.
We show here that the number of feedback loops in a network, as well as the eigenvalues of associated matrices, 
are determined by a structural property called trophic coherence, a measure of how neatly nodes fall into distinct levels.
Our theory correctly classifies a variety of networks -- including those derived from genes, metabolites, species,
neurons, words, computers and trading nations -- into two distinct regimes of high and low feedback,
and provides a null model to gauge the significance of related magnitudes.
Since trophic coherence suppresses feedback, whereas an absence of feedback alone does not lead to coherence, 
our work suggests that the reasons for `looplessness' in nature should be sought in coherence-inducing mechanisms.
\end{abstract}



%


\begin{center}
\end{center}

\begin{center}
\small{\url{http://www.pnas.org/content/early/2017/05/15/1613786114}}
\end{center}

\section*{Significance Statement}
Complex systems such as cells, brains or ecosystems are made up of many interconnected elements, each one acting on its neighbours, and sometimes influencing 
its own state via feedback loops. 
Certain biological networks have surprisingly few such loops. While this may be advantageous in various ways, it is not known how feedback is suppressed. 
We show that trophic coherence, a structural property of ecosystems, is key to the extent of feedback in these as well as in many other systems, including 
networks related to genes, neurons, metabolites, words, computers and trading nations. We derive mathematical expressions which provide a benchmark against which to 
examine empirical data, and conclude that `looplessness' in nature is probably a consequence of trophic coherence.

\section*{Introduction}

Feedback is a fundamental process in dynamical systems which occurs when the output of an element is coupled to its input.
In complex systems, this coupling can happen via feedback loops (or cycles) involving many elements, and hence the number and structure of such 
loops often determine important properties of the system as a whole \cite{May_qualitative}. 
In systems which can be represented as graphs, or networks, the combined effects of feedback loops are described by the 
spectrum of eigenvalues of the adjacency matrix (the matrix of ones and zeros representing the existence or absence of edges
between nodes) \cite{Arenas_synchro,brouwer2011spectra}. These eigenvalues 
can be related
to fundamental questions regarding both structure \cite{Fiedler,Newman_modularity} and dynamical processes --
including percolation \cite{Bollobas_percolation}, stability of dynamical elements \cite{May}, diffusion \cite{Barrat_book},
or synchronization of coupled oscillators \cite{Arenas_synchro}.
Feedback loops also play a role in the behaviour of many specific systems, such as 
robustness in gene regulatory networks \cite{zhang2012chaotic},
short-term memory in neural networks  \cite{johnson2013robust},
or systemic risk in financial networks \cite{dasgupta2014global}.

It has been observed that many biologically derived networks, such as food webs \cite{Milo824,Borrelli} and gene transcription networks \cite{BQS},
have far fewer feedback loops
than would be randomly expected,
or even none at all.
Given that acyclicity is the main requirement for being `qualitatively stable', or stable regardless of the details of 
dynamics \cite{May_qualitative},
one might suppose that this `loopless' architecture is an adaptation for stability or some other
functional advantage. However, in some cases it is not clear what the optimisation mechanism behind loop suppression might be.
In an ecosystem, for instance, how would a feedback cycle be eliminated if it happened to benefit the particular organisms involved?

It has recently been shown that the high linear stability of food webs
is determined mainly by a structural feature called `trophic coherence', a measure of how neatly nodes fall into 
distinct levels \cite{Johnson_trophic}.
Trophic coherence, moreover, has been found to play an important role in other structural and dynamical properties of 
networks \cite{klaise2016neurons,dominguez2016intervality}.
In order to investigate the relationship between trophic coherence and feedback, we here
define the `coherence ensemble' of graphs,
and obtain expressions for various magnitudes relating to the cycle structure and spectrum of eigenvalues of coherent 
but otherwise random networks.
We find that the number of cycles of length $\nu$ in a network can either grow or decay exponentially with $\nu$,
according to a `loop exponent', $\tau$, which is a function of trophic coherence. A corollary is that the expectation for the leading eigenvalue is  $\overline{\lambda_1}=e^\tau$.
Thus, depending on the sign of $\tau$ and hence on trophic coherence, a network can belong either to a `loopful' regime characterised by many cycles and high leading eigenvalues; or a `loopless' one in which cycles become scarcer with length, and all eigenvalues have real parts close to zero. In the loopless regime, the probability of drawing a directed acyclic graph tends to one with decreasing $\tau$.
We analyse a collection of empirically derived networks of several kinds, and find that they conform to our theoretical predictions, with
those networks with negative loop exponents displaying very few or no cycles.
The observation of scarcity of feedback in certain natural systems is therefore unsurprising, given their trophic coherence.

Our work also suggests the question: what mechanisms explain trophic coherence?
In the case of food webs, there are probably evolutionary pressures leading predators to specialise on prey 
on a narrow range of trophic levels \cite{Johnson_trophic}.
However, further research is needed to reveal other coherence-inducing mechanisms.

\section*{Results}

The results we present here are for graph ensembles; that is, we make statements about expected values or probability distributions over 
the sets of all possible graphs which meet certain constraints.
As we shall see, these results can shed light on the relationships between the structural features of many real-world networks,
to the extent that they can be regarded as random draws from a particular ensemble.
Paul Erd\H{o}s and Alfred R\'enyi pioneered this approach to graph theory with their analysis of the ensemble of all 
graphs with a given number of nodes $N$ and edges $L$ \cite{erdos1959random}.
The Erd\H{o}s-R\'enyi ensemble has two regimes, one for $L>N$ in which almost all graphs will include a giant connected component,
and another at $L<N$ in which no component will have more than $O(\ln N)$ nodes. 
Sometimes other characteristics are important. For example,
while node degrees (i.e. numbers of neighbours) in the Erd\H{o}s-R\'enyi ensemble are Poisson distributed,
real networks often have heavy-tailed degree distributions --
a property which affects many other topological features \cite{Newman_rev}.
Such systems might thus be better studied by means of the {\it configuration ensemble}, the set of all graphs with 
not only given $N$ and $L$, but also a given degree sequence \cite{molloy1995critical}.

In the same spirit, we here define the {\it coherence ensemble} as the set of directed graphs with a given number of nodes and degree sequence,
plus a specified trophic coherence.
We go on to show that in this ensemble there are also two regimes, depending on a single parameter, $\tau$, called the loop-exponent:
\begin{equation}
\tau=\ln \alpha + \frac{1}{2\tilde{q}^2}-\frac{1}{2q^2},
\label{eq_tau}
\end{equation}
where the branching factor $\alpha$ depends on the degree sequence, and $q$ and $\tilde{q}$ capture the trophic coherence
of a given network and that of its random expectation, respectively.
The coherence ensemble expectations for magnitudes such as the number of cycles of given length, or the leading eigenvalue, 
depend on $\tau$ exponentially. Therefore, as we shall go on to show, the sign of $\tau$ determines whether a network belongs 
to the ``loopless'' ($\tau<0$) or the ``loopful'' ($\tau>0$) regimes.

\paragraph{Definitions.} Consider the directed, unweighted graph given by 
the $N\times N$ adjacency matrix $A=(a_{ij})$, which has $L=\sum_{ij}a_{ij}$ directed edges.
The in- and out-degrees of node $i$ are $k_i^{in}=\sum_j a_{ij}$ and
$k_i^{out}=\sum_j a_{ji}$, respectively, and the mean degree is $\langle k\rangle=L/N$
(we shall use the notation $\langle \cdot\rangle$ to refer to averages over nodes in a given graph, as opposed to ensemble averages).
Note that the mean degree can be regarded as either the mean in-degree or the mean out-degree, since these coincide:
$\langle k\rangle=N^{-1}\sum_i k_i^{in}=N^{-1}\sum_i k_i^{out}$.
An important magnitude which depends only on degrees is the branching factor:
\begin{equation}
\alpha=\frac{\langle k^{in}k^{out}\rangle}{\langle k\rangle}.
\label{eq_alpha}
\end{equation}
Note that this magnitude, which together with trophic coherence determines the loop-exponent $\tau$, 
only depends on the mean degree $\langle k\rangle$ and the correlations between in- and out-degrees,
and not on other aspects of the degree distributions.

The eigenspectrum of $A$ is $\lbrace \lambda_i\rbrace$.
The trace of the $n$-th power of any square matrix $A$ can be expressed in terms of its eigenvalues 
as $\mbox{Tr}(A^n)=\sum_i \lambda_i^n$ \cite{CohnAlgebra}.
Therefore,
the distribution of eigenvalues, $p(\lambda)$, is related to powers of $A$ via its moments:
\begin{equation}
\langle\lambda^\nu\rangle = \frac{1}{N}\mbox{Tr}(A^\nu).
\label{eq_mom_trace}
\end{equation}
Since $A$ is not, in general, symmetric, its eigenvalues will be complex.
The trace of $A$ is real and invariant with respect to a change of basis,
so the eigenvalues of $A$ will always be distributed symmetrically around
the real axis \cite{CohnAlgebra}. Of particular interest is the eigenvalue with largest real part, $\lambda_1$ --
usually referred to as $A$'s leading eigenvalue.

A `basal node' is one with in-degree equal to zero.  
If a graph has at least one basal node (our assumption throughout), 
and every node belongs to at least one directed path 
which includes a basal node,
we can define the trophic level of each node $i$ as
\begin{equation}
s_i=1+\frac{1}{k_i^{in}}\sum_j a_{ij} s_j.
\label{eq_s_def}
\end{equation}
With no loss of generality for subsequent results, we define the trophic level
of basal nodes as $s_i=1$ ($\forall i$ such that $k_i^{in}=0$) \cite{Levine_levels}.
This is the convention in ecology, where the trophic level of a species informs as to its ecological function:
typically, plants have $s=1$, herbivores $s=2$, and omnivores and 
carnivores $s>2$.\footnote{Eq. (\ref{eq_s_def}) is similar to the definition of the PageRank 
algorithm used by the search engine Google \cite{brin1998anatomy}.
The main difference is that the sum in Eq. (\ref{eq_s_def}) is normalised by $k_i^{in}$, whereas PageRank divides
each term in the sum by $k_j^{out}$. Also, the small ``teleportation'' additive term which PageRank includes
to avoid problems with cycles is here the ``$+1$'' term which induces the hierarchy of trophic levels.
Both measures are related to diffusion processes; but while PageRank provides the probability of a node being reached 
by a ``random surfer'' (a random walker with some chance of teleportation), Eq. (\ref{eq_s_def}) provides a measure of 
how far the biomass arriving at a given node has travelled from the basal nodes.}
Note that Eq. (\ref{eq_s_def}) is a system of linear equations which can be solved whenever 
every node is on a path which begins at a basal node \cite{Johnson_trophic}.
Hence, despite the recurrent nature of this definition of trophic levels, the presence of cycles does not pose a problem.

In Ref. \cite{Johnson_trophic} we defined the `trophic difference' associated to each edge: $x_{ij}=s_{i}-s_{j}$.
The distribution of trophic differences over edges, $p(x)$, has mean 
$L^{-1}\sum_{ij}a_{ij}x_{ij}=1$
by definition,\footnote{This can be easily seen by noting that, for any node $i$, 
the average difference over its incoming edges is $\sum_{ij}a_{ij}(s_i-s_j)/k_i^{in}=1$.}
and we can measure the graph's `trophic coherence' with its standard deviation:
\begin{equation}
q=\sqrt{\frac{1}{L}\sum_{ij}a_{ij}x_{ij}^2 -1}.
\end{equation}
A graph will be more trophically coherent the closer $q$ is to zero, so we refer to $q$ as an 
`incoherence parameter'.
Maximal coherence, $q=0$, corresponds to a ``layered'' network in which every node has an integer trophic level,
and, as $q$ increases, the further the system departs from this ordered configuration \cite{Johnson_trophic,klaise2016neurons}.

The number of directed paths (henceforth `paths') of length $\nu$ in $A$ is $n_{\nu}=\sum_{ij}(A^\nu)_{ij}$. The number of directed cycles
(henceforth `cycles') of length $\nu$ is $m_{\nu}=\mbox{Tr}(A^\nu)$, which, according to Eq. (\ref{eq_mom_trace}), 
can be expressed as
$m_{\nu}=N\langle \lambda^\nu \rangle$. (Note that we are not referring here to simple paths and simple cycles, in which
no node can be repeated.) This definition of $m_\nu$ counts every unique cycle $\nu$ times,
so the number of unique cycles will be $m^u_\nu=m_\nu/\nu$.

The `directed configuration ensemble' is the set of all possible graphs with given
in- and out-degree sequences \cite{Charo}.
If the number of basal edges connected to basal nodes in a graph drawn from this ensemble is $L_B$, 
then for any node $i$ the expected proportion of in-neighbours
connected to a basal node will be $k^{in}_i L_B/L$. 
In order to obtain several expectations related to trophic coherence exactly, we define a modified version of this ensemble called 
the `basal ensemble', which is the subset of graphs from the directed configuration ensemble which satisfy the constraint that
the proportion of neighbours connected to basal nodes is exactly $k^{in}_i L_B/L$ for every node $i$.
It is straightforward to determine that in this ensemble the expectations for the trophic coherence and for the branching factor are given, 
respectively, by
\begin{equation}
\tilde{q}=\sqrt{\frac{L}{L_B}-1}
\label{eq_qcm}
\end{equation}
and
\begin{equation}
 \tilde{\alpha}=\frac{L-L_B}{N-B}
\label{eq_alpha_tilde}
\end{equation}
(where we use the notation $E(z)=\tilde{z}$ to refer to the expectation of magnitude $z$ in the basal ensemble).
The full derivation of these results can be found in SI Appendix.
In the limit $N\rightarrow \infty$, with $L/N\rightarrow \infty$, expectations in the basal ensemble and the directed configuration
ensemble converge. For finite graphs, we show numerically in SI Appendix that expectations in the two ensembles are close.
Equations (\ref{eq_qcm}) and (\ref{eq_alpha_tilde}) can therefore be considered reasonable null expectations for real networks
given only $N$, $L$, $B$ and $L_B$ -- i.e. in the absence of information regarding in-out-degree correlations or trophic coherence.

\paragraph{The coherence ensemble.}
Let us now consider the ensemble of directed
graphs which not only have given in- and out-degree distributions (as in the directed
configuration ensemble), but also given trophic coherence.
We shall refer to this as the `coherence ensemble',
and use the notation $E(z)=\overline{z}$ for
the expected values of quantities $z$ in this ensemble.
For networks in the coherence ensemble,
the probability of a randomly chosen path of length $\nu$ being a cycle can be obtained by considering
a random walk along the edges of the graph and computing the probability that it returns to the initial node after $\nu$ hops.
This constraint implies that the sum of the trophic differences $x_{k}$ over the $k=1,...\nu$ edges involved, $S=\sum_k x_k$, must be zero.
Let us approximate the differences $x_{k}$ as independent random variables drawn from the 
trophic difference
distribution $p(x)$.
According to the central limit theorem,
the distribution $p(S)$ will tend, with increasing $\nu$, to a Gaussian
with mean $\nu \langle x\rangle =\nu$ and variance $\nu q^2$.
Since cycles are paths which satisfy $S=0$, the expected proportion of paths of length $\nu$ that are cycles,
$\overline{c}_{\nu}$, will be proportional to $p(S=0)$. That is,
\begin{equation}
\overline{c}_{\nu} = B_\nu \frac{1}{\sqrt{\nu}q} \exp\left(-\frac{\nu}{2q^2}\right).
\label{eq_cnu}
\end{equation}
Not all the paths satisfying $S=0$ will return to the initial node, and this effect is accounted for by the factor $B_\nu$.
We can obtain $B_\nu$ by particularising for the basal ensemble case, for which $q$ 
is given by Eq. (\ref{eq_qcm}), and  $c_\nu=\tilde{\alpha}/L$ 
(see SI Appendix).
Inserting these values into Eq. (\ref{eq_cnu}), we have
\begin{equation}
B_\nu=\frac{\tilde{\alpha}}{L}\sqrt{\nu}\tilde{q}\exp\left(\frac{\nu}{2\tilde{q}^2} \right).
\end{equation}
Therefore, an approximate expression for $\overline{c}_\nu$ is
\begin{equation}
 \overline{c}_\nu=\frac{\tilde{\alpha}}{L} \frac{\tilde{q}}{q} \exp\left[\frac{\nu}{2}\left(\frac{1}{\tilde{q}^2}-\frac{1}{q^2}\right) \right].
\label{eq_cnuq}
 \end{equation}
The expected proportion of paths of size $\nu$ which are cycles can
thus 
either decrease or increase
exponentially with $\nu$, depending on whether a particular graph is more or less trophically coherent than the null expectation
given its degree sequence.
Eq. (\ref{eq_cnuq}) was obtained using the central limit theorem, and so should only be valid for moderately large $\nu$. However,
if the distribution of differences, $p(x)$, is approximately normal, it will be a good approximation also at low values of $\nu$.
We have also assumed the trophic differences of each path to be independent random variables 
drawn from $p(x)$,
an approximation which will hold as long as there are no significant correlations between these differences.

Let us now assume that the total number of paths in the coherence ensemble is given approximately by 
$\overline{n}_\nu\simeq L\alpha^{\nu-1}$, as in the basal and the directed configuration ensembles
-- i.e. irrespective of $q$ (see SI Appendix).
This is a reasonable assumption, at least for low $\nu$, since
$\alpha$ is the key element determining
the number of ways a set of edges can be concatenated. (For finite $N$, the approximation may break down at high $\nu$ and low $q$,
because the maximum path length will be shorter in highly coherent graphs than in random ones.)
Combining this with Eq. (\ref{eq_cnuq}) 
we obtain the expected number of cycles of length $\nu$:
\begin{equation}
\overline{m}_\nu=\frac{\tilde{\alpha}\tilde{q}}{\alpha q} e^{\tau \nu},
\label{eq_m}
\end{equation}
where the `loop exponent' $\tau$ has already been supplied in Eq. (\ref{eq_tau}).
The term $1/\tilde{q}^2-1/q^2$ in Eq. (\ref{eq_tau})
will be negative for networks which are more coherent than the random expectation ($q<\tilde{q}$),
and positive for those which are less so; while the sign of $\ln \alpha$ depends on whether $\alpha$ is greater or less than $1$.
Eq. (\ref{eq_m}) implies
that the expected number of cycles of length $\nu$ in a graph can either grow
exponentially with $\nu$, when $\tau>0$; or decrease exponentially, if $\tau<0$. Thus, which of these two regimes a given graph
finds itself in is determined by
the correlation between in- and out-degrees,
$\alpha=\langle k^{in}k^{out}\rangle/\langle k\rangle$;
the proportion of
edges which connect to basal nodes, $L_B/L$ (via $\tilde{q}=\sqrt{L/L_B-1}$);
and the trophic coherence, given by $q$.
Note that, as mentioned above, the definition of $m_\nu$ counts each cycle $\nu$ times, so the expected number of unique cycles is
\begin{equation}
\overline{m}^u_\nu=\frac{\tilde{\alpha}\tilde{q}}{\alpha q}\frac{e^{\tau \nu}}{\nu}.
\label{eq_munique}
\end{equation}

The number of cycles is related to the eigenspectrum of the adjacency matrix through Eq. (\ref{eq_mom_trace}).
Therefore, from Eq. (\ref{eq_m}) we have
that the expected value of the $\nu$-th moment of the distribution of
eigenvalues is
\begin{equation}
\overline{\langle \lambda^\nu\rangle} = \frac{1}{N}\sum_i \overline{\lambda_i}^\nu = \frac{1}{N}\frac{\tilde{\alpha}\tilde{q}}{\alpha q}e^{\tau \nu}.
\label{eq_sum}
\end{equation}
We can use this relation to obtain, for the coherence ensemble,
the expected value of the leading eigenvalue by considering the limit of large $\nu$:
\begin{equation}
\lim_{\nu \to +\infty} \left(\sum_i \overline{\lambda_i}^\nu\right)^\frac{1}{\nu}=\overline{\lambda_1}=e^{\tau}.
\label{eq_lambda_1}
\end{equation}

The expressions for the configuration ensemble can be recovered by taking $q=\tilde{q}$,
 which, according to Eq. (\ref{eq_tau}),
 implies $\tau=\ln \alpha$. Thus, the leading eigenvalue in the directed configuration ensemble is
$\tilde{\lambda}_1=\alpha=\langle k^{in}k^{out}\rangle/\langle k\rangle$.
If the graph were symmetric ($k_i^{in}=k_i^{out}=k_i$, $\forall i$), we would have
$\tilde{\lambda}_1^{Sym}=\langle k^2\rangle/\langle k\rangle$; while for the Erd\H{o}s-R\'enyi ensemble we obtain $\tilde{\lambda}_1^{ER}=1+\langle k\rangle$.
These particular cases are in agreement with previous mean-field results for these ensembles \cite{Chung_PNAS,Newman_PRE}.
The expected distribution of eigenvalues is entirely defined by its full set of moments,
as given by Eq. (\ref{eq_sum}). For instance, the moment-generating function for graphs with given $\tau$ is
\begin{equation}
 M_\lambda(t)=\sum_{\nu=0}^\infty\frac{t^\nu}{\nu!}\overline{\langle \lambda^\nu\rangle}=
\left(1-\frac{1}{N}\frac{\tilde{\alpha}\tilde{q}}{\alpha q}\right)+\frac{1}{N}\frac{\tilde{\alpha}\tilde{q}}{\alpha q}\exp\left(t e^{\tau}\right).
\label{eq_MGF}
\end{equation}

\begin{figure}
\centering
\includegraphics[width=1\linewidth]{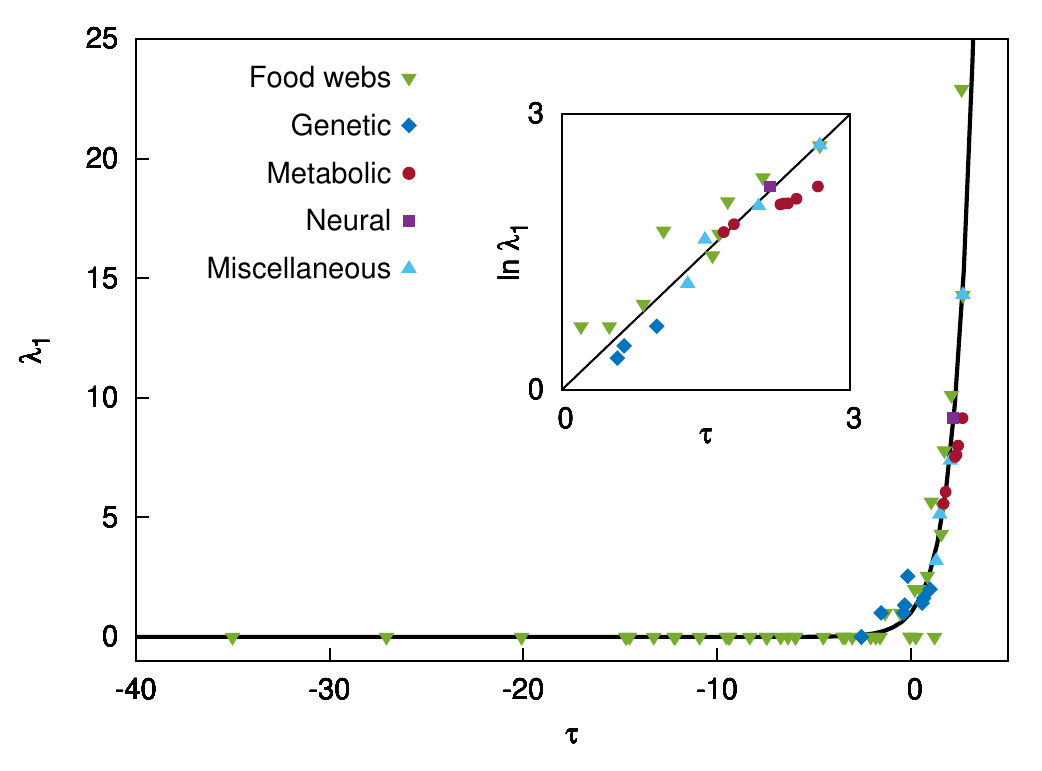}
\caption{
Loop exponents $\tau$ are informative of leading eigenvalues for a variety of empirical networks.
Here we plot leading eigenvalues $\lambda_1$ of several directed networks, against $\tau$ as given by Eq. (\ref{eq_tau});
symbols indicate graph data derived from
food webs (green down-pointing triangles), gene regulatory networks (dark blue diamonds), metabolic networks 
(burgundy circles), a neural network (purple square), and other miscellaneous networks (light blue up-pointing triangles). 
Line: Expected leading eigenvalue $\overline{\lambda_1}$ in the coherence ensemble, as given by Eq. (\ref{eq_lambda_1}).
Inset: Semi-log version of the positive quadrant of the main panel (Pearson's correlation coefficient: $r^2=0.87$).
For these results, self-edges were removed from the networks; a similar figure in which self-edges are included can be found in SI.
Details for each network, including references, are listed in the tables of SI Appendix.
}
\label{fig_tau}
\end{figure}

\paragraph{Empirical networks.}
We have obtained the adjacency matrices of $62$ directed networks from various sources. Several details of each, including references, 
are listed in Tables S1--S4 of SI Appendix.
There are three broad classes of biologically derived network in our data set: food webs, gene regulatory networks, and metabolic networks.
We also include a neural network, and several man-made networks: two of international trade, a P2P file-sharing network, and a network of 
concatenated English words. In all cases we have removed self-edges if present, mainly because these are not reported for many of the networks,
and the nature of self-interaction is often different from that occurring between elements. However, in 
SI Appendix
we also analyse the same networks while conserving self-edges when reported, and the results do not differ significantly.
Figure \ref{fig_tau} displays the leading eigenvalues, $\lambda_1$, against $\tau$ for 
the $62$ networks, with different classes of network identified by the symbols, as indicated.
The coherence ensemble expected value given by Eq. (\ref{eq_tau}), shown with a line, provides a good estimate of almost all the 
empirical values.
The inset shows the positive quadrant on a semi-log scale.
Of the 62 networks in our data set, 36 have $\tau<0$ and 26 have $\tau>0$. The mean values of $\lambda_1$ for these are, respectively, 
$\lambda_1(\tau<0)=0.22\pm0.54$ and $\lambda_1(\tau>0)=6.1\pm7.4$.
In other words, the two regimes are separated by an order 
of magnitude in the leading eigenvalue.

Table \ref{table_comparison} shows the mean and standard deviation of several magnitudes for the three main classes of biologically-derived network 
in our data set. The first three rows are for the ratios of measured values to the basal ensemble expectations. 
The 
graphs corresponding to
food webs are significantly coherent ($q/\tilde{q}<1$) and have slightly lower mean trophic levels than the expectation ($\langle s\rangle/\tilde{s}\lesssim 1$).
The networks derived from gene regulation have coherence and mean trophic levels which are very close to their expected values. Meanwhile, the 
networks linked to metabolism are significantly incoherent ($q/\tilde{q}>1$) and have mean trophic levels which are higher than expected ($\langle s\rangle/\tilde{s}>1$).
The measured values of $\alpha$ are in all three classes slightly higher than the expectation, but in the cases of food webs and gene regulatory networks, the difference 
is within one standard deviation. However, the metabolism-related networks display marked positive correlations between in- and out-degrees ($\alpha/\tilde{\alpha}>1$).
The fifth row shows the proportion of networks in each class which have negative $\tau$. For the food web and gene regulatory network 
data, it is $74\%$ and $63\%$, respectively, while the metabolism-related
networks are all in the positive $\tau$ regime.
This leads to average leading eigenvalues, shown in the fourth row, which are much greater for metabolic network 
data than for food webs or gene regulatory-related
networks. 
The sixth row gives the proportion of networks in each class which are acyclic, a feature we discuss in the next section.
In SI Appendix we show an example of each kind of network to illustrate the wide variety of trophic structures found among natural systems.

\begin{table}
\centering
\caption{
Mean values and standard deviation of the ratios $q/\tilde{q}$, $\langle s\rangle/\tilde{s}$
and $\alpha/\tilde{\alpha}$, and of the leading eigenvalue $\lambda_1$,
for three classes of biologically-derived network; and fractions of these networks to have $\tau<0$, and to be acyclic, of the total in our data set.}
 \begin{tabular}{@{}lcccc}
 & Food webs & Genetic & Metabolic \\
 \hline
 \hline 
 $q/\tilde{q}$ & $0.44 \pm 0.17$ & $0.99 \pm 0.05$ & $1.81\pm 0.11$
 \\
 $\langle s\rangle/\tilde{s}$ & $0.88 \pm 0.18$ &  $1.00 \pm 0.001$ & $ 2.05 \pm 0.01$
\\
$\alpha/\tilde{\alpha}$ & $1.02 \pm 0.23$ &  $1.19 \pm 0.34$ & $3.98 \pm 1.04$
\\
$\lambda_1$ & $1.54 \pm 4.09$ & $1.36 \pm 0.75$ & $7.36 \pm 1.20$ 
\\
$\tau<0$ & $31/42$ & $5/8$ & $0/7$
\\
Acyclic & $31/42$ & $1/8$ & $0/7$
\\
\label{table_comparison}
 \end{tabular}
\end{table}

\paragraph{Directed acyclic graphs.}
Let us consider the probability that a graph randomly chosen from the coherence ensemble
will have exactly $m_\nu$ cycles of length $\nu$.
We shall assume that each path is an
independent random event with two possible outcomes: with
probability $\overline{c}_\nu$ it is a cycle, while with $1-\overline{c}_\nu$ it is not.
The number of cycles $m_\nu$ will therefore be binomially distributed:
\begin{equation}
p(m_\nu)={\tilde{n}_\nu \choose m_\nu}\overline{c}_\nu^{m_\nu} (1-\overline{c}_\nu)^{\tilde{n}_\nu-m_\nu}.
\label{eq_binom}
\end{equation}
We can use this distribution to obtain the probability that a network from the coherence ensemble would have no directed 
cycles of length greater or equal to $n$:
\begin{equation}
P_n=\prod_{\nu=n}^{\infty}p(m_\nu=0).
\end{equation}
For instance, the probability that a network drawn randomly from this ensemble would be acyclic is
\begin{eqnarray}
P_{acyclic}=\prod_{\nu=2}^{\infty}p(m_\nu=0)=\\ \nonumber
\prod_{\nu=2}^{\infty}\left\lbrace 1-\frac{\tilde{\alpha}}{L}\frac{\tilde{q}}{q} \exp\left[ \frac{\nu}{2}\left(\frac{1}{\tilde{q}^2}-\frac{1}{q^2}\right)\right] \right\rbrace^{L\alpha^{\nu-1}}.
\nonumber
\end{eqnarray}

Taking logarithms and considering graphs with sufficiently negative $\tau$ that we can use the approximation
$\ln(1-x)\simeq -x$, we have
\begin{equation}
\ln P_{acyclic}\simeq -\frac{\tilde{\alpha}\tilde{q}}{\alpha q} \sum_{\nu=1}^\infty e^{\tau \nu};
\end{equation}
and, after performing the sum and some algebra,
\begin{equation}
P_{acyclic}\simeq \exp\left[-\frac{\tilde{\alpha}\tilde{q}}{\alpha q}\frac{1}{(e^{-\tau}-1)} \right].
\label{P_DAG}
\end{equation}
Therefore, as $\tau\rightarrow-\infty$, networks are almost always acyclic.
We note that these sums include small values of $\nu$ for which results are only approximate unless the distribution of trophic differences,
$p(x)$, is Gaussian.

\begin{figure}[t!]
\begin{center}
\includegraphics[width=1\linewidth]{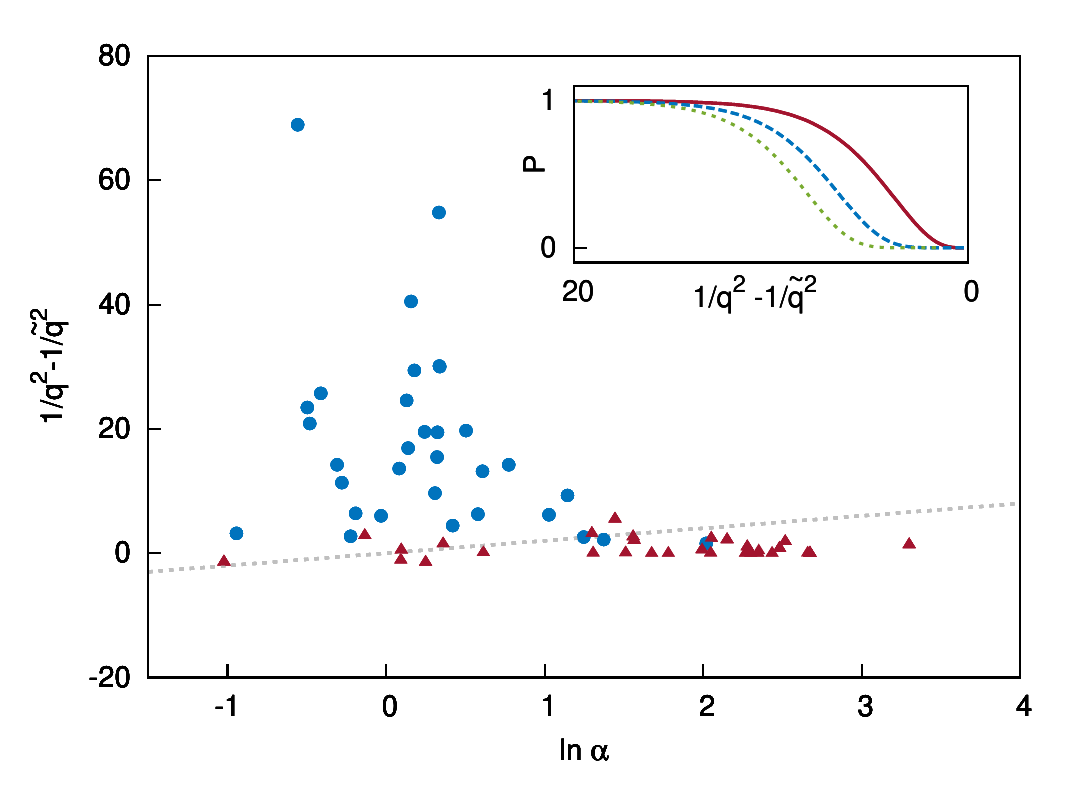}
\end{center}
\caption{
The components of the loop exponent $\tau$, coherence vs in-degree-out-degree correlations, are predictive of whether empirical networks will have cycles.
Here we show a scatter plot of several networks according to 
$1/q^2-1/\tilde{q}^{2}$ and $\ln\alpha$, the two terms in Eq. (\ref{eq_tau}).
Blue circles: networks with no cycles. Burgundy triangles: networks with at least one cycle.
The $\tau=0$ line is shown with a dashed line (the negative $\tau$ regime falls above the line).
Details for each network, including references, are listed in the tables of SI Appendix.
Inset: Probabilities of networks in the coherence ensemble being acyclic, according to Eq. (\ref{P_DAG}), 
as a function of $1/q^2-1/\tilde{q}^{2}$. Solid line: $L/L_B=10$ and $\alpha=1$; dashed line: $L/L_B=100$ and $\alpha=1$;
dotted line: $L/L_B=100$ and $\alpha=2$.
}
\label{fig_alpha}
\end{figure}

Figure \ref{fig_alpha} is a scatter plot of our set of empirical networks according to the terms in Eq. (\ref{eq_tau}):
$1/q^2-1/\tilde{q}^{2}$ and $\ln\alpha$.
Each network is represented with either a triangle or a circle depending on whether it has cycles or not, respectively.
The curve $\tau=0$ separates the two regimes,
and it is clear that while almost all the networks in the positive $\tau$ regime have cycles (the exceptions being 
two food webs), as one moves into the negative $\tau$ regime most of the examples are acyclic.
The inset shows the probability of a network randomly drawn from the coherence ensemble being acyclic, 
as given by Eq. (\ref{P_DAG}) and indicated in the caption.
One can compute, for each empirical network, the probability that it is acyclic according to Eq. (\ref{P_DAG}).
Thus, given only a network's degree sequence and trophic coherence, 
we would expect it to be acyclic if $P_{acyclic}>0.5$
(note that a network might be in the $\tau<0$ regime yet still be predicted to have cycles by this decision rule).
We find that, out of the 62 networks, eight are incorrectly classified: seven food webs are acyclic despite being 
predicted to have cycles, and one gene regulatory network would be expected (by a small margin) to be acyclic but is not.
The prediction is therefore accurate in $87\%$ of instances.

\section*{Discussion}
We have shown that a directed network can belong to either of two regimes characterised by fundamentally different cycle structures,
depending on the sign of a single parameter, 
$\tau$, which is a function of the trophic coherence and the 
branching factor, as given by Eq. (\ref{eq_tau}).
Since the expected number of cycles of length $\nu$ is proportional to $e^{\tau\nu}$, positive $\tau$ implies an exponentially growing
number of cycles with length, while for
negative $\tau$ the probability of finding cycles is vanishing.
This, in turn, has a crucial effect on the spectral properties of graphs: in particular, the expected value of the
leading eigenvalue of the adjacency matrix is $\overline{\lambda_1}=e^{\tau}$. A corollary is that graphs drawn randomly
from the negative $\tau$ regime have a high probability of being directed acyclic graphs, the main requisite for
qualitative stability \cite{May_qualitative}.

Our results provide expected values for what we have called the coherence ensemble -- the set of directed graphs
with a given degree sequence and trophic coherence -- and do not, therefore, place bounds on the possible values a given network can exhibit.
However, analysis of a set of empirically-derived networks of various kinds shows that in most cases these expected values are very good 
approximations to the ones measured, suggesting that the coherence ensemble may be an appropriate null model to use in many cases.
We should note also that we have focused on binary networks (those with adjacency matrices of only ones and zeros). While some of the results
could be extended to weighted networks in a straightforward way, it is not so obvious how concepts such as trophic coherence should be understood 
when a distinction between excitatory and inhibitory interactions is made. We leave such questions for future work.

The fact that many biologically-derived networks have surprisingly few 
feedback loops
has recently been attributed
to considerations of robustness \cite{BQS}, stability \cite{Borrelli}, and to an ``inherent directionality'' \cite{Vir_loops}.
Our results are compatible with the latter, since network directionality would ensue from trophic coherence
(that is, since the distribution of differences $p(x)$ is centred at $1$ and has variance $q^2$, the
expected number of edges with $x<0$ is $L\int_{-\infty}^0 p(x)dx$, a monotonically increasing function of $q$).
However,
neither a
suppression of cycles
nor an imposed directionality
will in themeselves
induce trophic coherence, as can be easily seen in the case of the 
``cascade model'' \cite{Cohen_1}. In this network assembly model, 
there is a
strict
hierarchy of nodes; directed edges are placed at random with the sole constraint that 
the out-node must be below the in-node in the hierarchy, thus emulating the situation in many food webs where predators 
tend to be larger than their prey.
Such networks are by construction acyclic
and directional,
yet they do not exhibit significant trophic coherence \cite{Johnson_trophic}.
On the other hand, our analysis indicates that any network formation processes which tended to induce a certain trophic coherence 
would confer the properties of a low or negative $\tau$ on a system, without
necessarily being the result of an optimization for low feedback.
For example, in ecosystems many features of species, such as body size and metabolic rate, are related to trophic levels.
Since predators often specialise in consuming prey with specific characteristics, they naturally focus on relatively narrow 
trophic ranges,
a mechanism which could lead to networks that are more coherent than the random expectation.
This idea is borne out by generative network models which capture this effect --
namely, the `preferential preying model' presented in Ref. \cite{Johnson_trophic}
(which produces acyclic graphs with tunable trophic coherence)
and an extension of this model studied in Ref. \cite{klaise2016neurons}
(which can set the trophic coherence of graphs with cycles).
However, relatively little is yet known about the mechanisms which might lead to trophic coherence more generally.

While we have argued here that looplessness should be regarded as an effect of trophic coherence, this naturally
moves the challenge to establishing the origins of trophic coherence.
Further research is needed to address this issue, possibly involving the relation between trophic levels 
and the functional roles of nodes.
This view has interesting parallels with recent work on node roles in generic directed networks, 
based on topological similarity, which when applied to food webs reveals trophic structure \cite{cooper2010role,beguerisse2014interest}.
Functional groups have also been uncovered in ecosystems using stochastic block models,
which can take non-trophic interactions into account
\cite{kefi2015network,kefi2016structured}.

A relation between node function and trophic level may exist in systems other than ecological ones. For instance, 
in the word adjacency network of the children's book {\it Green Eggs and Ham}, by Dr Seuss, we find 
that the mean trophic level of common nouns is $s_{noun}=1.4\pm1.2$, while that of verbs is $s_{verb}=7.0\pm2.7$
(see Fig. S4 in SI Appendix).
This shows that in networks where node function is encoded in trophic levels, any mechanism whereby edges tended to occur between 
nodes with specific functions might develop non-trivial coherence (or incoherence).
More broadly, it also
suggests that the trophic structure of directed networks may provide insights into their function and dynamics.
Classifying
nodes by trophic level, as has long been standard in ecology, might also tell us something about the functions of, say, genes, metabolites, 
neurons, economic agents, or words in unknown languages.
In view of these considerations, 
we believe that further exploration of the trophic structure of networks, and its relation to function and dynamics, will
prove a fruitful avenue for learning about many complex systems.








\section*{Acknowledgements}
Many thanks to M.A. Mu\~noz and V. Dom\'inguez-Garc\'ia for a fruitful collaboration from which some of these ideas sprouted.
We are grateful to I. Johnston, M. Ib\'a\~nez Berganza and J. Klaise for useful discussions.
Thanks also to L. Albergante, J. Dunne, U. Jacob, R.M. Thompson, C.R. Townsend,
U. Alon, M.E.J. Newman, W. de Nooy, and J. Leskovec for providing data or making them available online (see SI Appendix.)


\newpage


\begin{thebibliography}{10}

\bibitem{May_qualitative}
May RM (1973) Qualitative stability in model ecosystems.
\newblock {\em Ecology} 54:638--41.

\bibitem{Arenas_synchro}
Arenas A, D\'iaz-Guilera A, Kurths J, Moreno Y, Zhou C (2008) Synchronization
  in complex networks.
\newblock {\em Phys. Rep.} 469:93--153.

\bibitem{brouwer2011spectra}
Brouwer AE, Haemers WH (2011) {\em Spectra of graphs}.
\newblock (Springer Science \& Business Media).

\bibitem{Fiedler}
Fiedler M (1973) Algebraic connectivity of graphs.
\newblock {\em Czech. Math. J} 23:298--305.

\bibitem{Newman_modularity}
Newman MEJ (2006) Modularity and community structure in networks.
\newblock {\em Proc. Natl. Acad. Sci. USA} 103:8577--8582.

\bibitem{Bollobas_percolation}
Bollob\'as B, Borgs C, Chayes J, Riordan O (2010) Percolation on dense graph
  sequences.
\newblock {\em Ann. Prob.} 38:150--183.

\bibitem{May}
May RM (1972) Will a large complex system be stable.
\newblock {\em Nature} 238:413--14.

\bibitem{Barrat_book}
Barrat A, Barth\'elemy M, Vespignani A (2008) {\em Dynamical Processes on
  Complex Networks}.
\newblock (Cambridge University Press, Cambridge).

\bibitem{zhang2012chaotic}
Zhang Z et~al. (2012) Chaotic motifs in gene regulatory networks.
\newblock {\em PLOS ONE} 7(7):e39355.

\bibitem{johnson2013robust}
Johnson S, Marro J, Torres JJ (2013) Robust short-term memory without synaptic
  learning.
\newblock {\em PLOS ONE} 8(1):e50276.

\bibitem{dasgupta2014global}
DasGupta B, Kaligounder L (2014) On global stability of financial networks.
\newblock {\em Journal of Complex Networks} 2(3):313--354.

\bibitem{Milo824}
Milo R et~al. (2002) Network motifs: Simple building blocks of complex
  networks.
\newblock {\em Science} 298(5594):824--827.

\bibitem{Borrelli}
Borrelli JJ (2015) Selection against instability: stable subgraphs are most
  frequent in empirical food webs.
\newblock {\em Oikos}.

\bibitem{BQS}
Albergante L, Blow JJ, Newman TJ (2014) Buffered {Q}ualitative {S}tability
  explains the robustness and evolvability of transcriptional networks.
\newblock {\em eLife} 3:e02863.

\bibitem{Johnson_trophic}
Johnson S, Dom\'inguez-Garc\'ia V, Donetti L, Mu\~noz MA (2014) Trophic
  coherence determines food-web stability.
\newblock {\em Proc. Natl. Acad. Sci. USA} 111(50):17923--17928.

\bibitem{klaise2016neurons}
Klaise J, Johnson S (2016) From neurons to epidemics: How trophic coherence
  affects spreading processes.
\newblock {\em Chaos} 26(065310).

\bibitem{dominguez2016intervality}
Dom{\'\i}nguez-Garc{\'\i}a V, Johnson S, Mu{\~n}oz MA (2016) Intervality and
  coherence in complex networks.
\newblock {\em Chaos} 26(065308).

\bibitem{erdos1959random}
Erd\H{o}s P, R\'enyi A (1959) On random graphs {I}.
\newblock {\em Publ. Math. Debrecen} 6:290--297.

\bibitem{Newman_rev}
Newman MEJ (2003) The structure and function of complex networks.
\newblock {\em SIAM Review} 45:167--256.

\bibitem{molloy1995critical}
Molloy M, Reed B (1995) A critical point for random graphs with a given degree
  sequence.
\newblock {\em Random structures \& algorithms} 6(2-3):161--180.

\bibitem{CohnAlgebra}
Cohn PM (1982) {\em Algebra}.
\newblock (Wiley).

\bibitem{Levine_levels}
Levine S (1980) Several measures of trophic structure applicable to complex
  food webs.
\newblock {\em J. Theor. Biol.} 83:195--207.

\bibitem{brin1998anatomy}
Brin S, Page L (1998) The anatomy of a large-scale hypertextual web search
  engine.
\newblock {\em Computer networks and ISDN systems} 30(1):107--117.

\bibitem{Charo}
Kim H, {Del Genio} CI, Bassler KE, Toroczkai Z (2012) Constructing and sampling
  directed graphs with given degree sequences.
\newblock {\em New J. Phys.} 14:023012.

\bibitem{Chung_PNAS}
Chung F, Lu L, Vu V (2003) Spectra of random graphs with given expected
  degrees.
\newblock {\em Proc. Natl. Acad. Sci. USA} 100(11):6313--6318.

\bibitem{Newman_PRE}
Nadakuditi RR, Newman MEJ (2013) Spectra of random graphs with arbitrary
  expected degrees.
\newblock {\em Phys. Rev. E} 87:012803.

\bibitem{Vir_loops}
Dom\'inguez-Garc\'ia V, Pigolotti S, Mu{\~n}oz MA (2014) Inherent
  directionality explains the lack of feedback loops in empirical networks.
\newblock {\em Sci. Rep.} 4:7497.

\bibitem{Cohen_1}
Cohen JE, Newman CM (1985) A stochastic theory of community food webs {I}.
  models and aggregated data.
\newblock {\em Proc. R. Soc. London Ser. B.} 224:421--448.

\bibitem{cooper2010role}
Cooper K, Barahona M (2010) Role-based similarity in directed networks.
\newblock {\em arXiv preprint arXiv:1012.2726}.

\bibitem{beguerisse2014interest}
Beguerisse-D{\'\i}az M, Garduno-Hern{\'a}ndez G, Vangelov B, Yaliraki SN,
  Barahona M (2014) Interest communities and flow roles in directed networks:
  the twitter network of the uk riots.
\newblock {\em Journal of The Royal Society Interface} 11(101):20140940.

\bibitem{kefi2015network}
K{\'e}fi S et~al. (2015) Network structure beyond food webs: mapping
  non-trophic and trophic interactions on chilean rocky shores.
\newblock {\em Ecology} 96(1):291--303.

\bibitem{kefi2016structured}
K{\'e}fi S, Miele V, Wieters EA, Navarrete SA, Berlow EL (2016) How structured
  is the entangled bank? the surprisingly simple organization of multiplex
  ecological networks leads to increased persistence and resilience.
\newblock {\em PLoS Biol} 14(8):e1002527.

\end{thebibliography}

\end{document}


\maketitle

\vspace{2.5cm}

\section*{\underline{CONTENTS}}

\section{Graph ensembles}

\subsection{The directed configuration ensemble}

\subsection{The basal ensemble}

\subsection{Equivalence of ensembles}

\section{Networks with self-cycles}

\section{Network data}

\subsection{Green Eggs and Ham words network}

\newpage

\setcounter{section}{0}

\section{Graph ensembles}

\subsection{The directed configuration ensemble}

This is the set of all graphs with given sequences of in- and out-degrees \cite{Charo}.
Using
this ensemble as a null-model,
we can obtain the expected numbers of paths and cycles by inserting the expected
value of the adjacency matrix for large graphs, $\hat{a}_{ij}=k_{i}^{out}k_{j}^{in}/L$, in the above definitions (we shall use the notation
$\hat{z}$ to refer to the expected value of a magnitude $z$ in the directed configuration ensemble).
Thus, the expected
number of paths in this ensemble is
\begin{equation}
\hat{n}_{\nu}=L\alpha^{\nu-1},
\label{eq_ncm}
\end{equation}
while the expected number of those paths which are cycles is
\begin{equation}
\hat{m}_{\nu}=\alpha^{\nu},
\label{eq_mcm}
\end{equation}
where
\begin{equation}
\alpha=\frac{\langle k^{in}k^{out}\rangle}{\langle k\rangle}.
\label{eq_alpha}
\end{equation}
The branching factor $\alpha$
captures the correlation between the in- and out-degrees of nodes
(i.e. $\alpha>\langle k\rangle$ indicates a positive correlation, with high in-degree 
nodes also tending to have high out-degree, while 
$\alpha<\langle k\rangle$ means this correlation is negative).

\subsection{The basal ensemble}

Let us consider the ensemble
of random graphs which satisfy the following constraint: for every non-basal node, the proportion of
in-coming edges which connect to basal nodes is the same. This is a sufficient condition for all
non-basal nodes to have the same trophic level.
More formally, if a network in this ensemble has $B$ basal nodes, 
$N-B$ non-basal nodes, $L$ edges, and $L_B$ edges connecting to basal nodes,
then 
every non-basal node $i$ with in-degree $k_i^{in}$ receives $k_i^{in}L_B/L$ edges from basal nodes
and $(1-L_B/L) k_i^{in}$ from non-basal nodes.
Note that this constraint does not affect the expectations obtained above for the more general 
directed configuration ensemble, so the expected numbers of paths and cycles are, respectively, 
\begin{equation}
\tilde{n}_{\nu}=L\alpha^{\nu-1}
\label{eq_nbe}
\end{equation}
and
\begin{equation}
\tilde{m}_{\nu}=\alpha^{\nu}
\label{eq_mbe}
\end{equation}
(where we have used the notation $\tilde{z}$ for the expected value of a magnitude $z$ in the basal ensemble).
However, the fixed proportion of basal in-neighbours allows us also to derive
expected values for several magnitudes in this ensemble, given $\lbrace N, B, L, L_B\rbrace$, as follows.

We recall from the main text [Eq. (4)] 
that the trophic level of each node $i$ is
\begin{equation}
s_i=1+\frac{1}{k_i^{in}}\sum_j a_{ij} s_j
\label{eq_s_def}
\end{equation}
if is non-basal (i.e. $k_i^{in}>0$), and $s_i=1$ if $i$ is basal ($k_i^{in}=0$) \cite{Levine_levels}.
Let $\tilde{s}_{nb}$ be the expected trophic level in the basal ensemble of a node given that it is not basal.
Since the expected proportion 
of in-neighbours which are basal, for such a node, will be $L_B/L$, we have from 
Eq. (\ref{eq_s_def}) that
\begin{equation}
\tilde{s}_{nb}=1+\frac{L_B+(L-L_B)\tilde{s}_{nb}}{L},
\label{eq_snb}
\end{equation}
and so $\tilde{s}_{nb}=L/L_B+1$.
The mean trophic level of the network is therefore 
\begin{equation}
\tilde{s}=1+\left(1-\frac{B}{N}\right)\frac{L}{L_B}.
\label{eq_smed}
\end{equation}
In the basal ensemble
there will be two types of edges: those which emanate from basal nodes, with trophic difference
$x=\tilde{s}_{nb}-1$, and those between non-basal nodes, with $x=0$. In other words, the distribution of differences 
will be
\begin{equation}
p(x)=\frac{L_B}{L} \delta\left(x-\frac{L}{L_B}\right)+\left(1-\frac{L_B}{L}\right)\delta(x),
\label{eq_px_bimod}
\end{equation}
where $\delta$ is the Dirac delta function.
The trophic coherence associated with this distribution is
\begin{equation}
\tilde{q}=\sqrt{\frac{L}{L_B}-1}.
\label{eq_qcm_mm}
\end{equation}

We can also obtain the expected value of $\alpha$ in the basal ensemble. 
Every node with $k^{in}\neq 0$ (i.e. every non-basal node)
has in-degree $k^{in}=L/(N-B)$;
and while the out-degree of such a node is not determined, the expected value is
$\tilde{k}^{out}=(L-L_B)/(N-B)$.
Inserting these values into Eq. (\ref{eq_alpha}) yields
\begin{equation}
 \tilde{\alpha}=\frac{L-L_B}{N-B}.
\end{equation}
Note that this is the mean degree that would result if all basal nodes and edges emanating from basal nodes were eliminated
from the network.
As in the configuration ensemble, the expected proportion of paths of length $\nu$ which are cycles, in the 
basal ensemble, is
\begin{equation}
 \tilde{c}_\nu=\frac{\tilde{\alpha}}{L}.
 \label{eq_cmf}
\end{equation}

The basal ensemble does not include any structure which might lead to variance in the trophic levels of 
non-basal nodes, and so the bimodal $p(x)$ given by Eq. (\ref{eq_px_bimod}) is exact.
For networks in which different non-basal nodes are connected to differing proportions of basal nodes,
this expression will be a valid approximation
only when the separation of the two modes of $p(x)$
is much larger than the spread about them.

%
%

\subsection{Equivalence of ensembles}

Let $\kappa_i$ be the proportion of in-coming edges to node $i$ which emanate from a
basal node. In the directed configuration ensemble, the expectation for $\kappa_i$
is $\hat{\kappa}_i=L_B/L$, $\forall i$. The basal ensemble is the subset of graphs
from the directed configuration ensemble which satisfy $\kappa_i=L_B/L$, $\forall i$,
exactly, not just in expectation.
Thanks to this constraint, all non-basal nodes have the same trophic level, given by
Eq. (\ref{eq_snb}), in the basal ensemble. For finite graphs drawn from the directed
configuration ensemble, there will be some variation in the trophic levels of non-basal
nodes, leading to discrepancies with respect to the expectations of quantities such as
$q$, $\alpha$ and $\langle s\rangle$.
However, in the limit $N\rightarrow \infty$, with $L/N\rightarrow \infty$, 
expectations in the basal ensemble and the directed configuration ensemble converge,
since $\kappa_i\rightarrow \hat{\kappa}_i$, $\forall i$. 
This suggests that the basal ensemble might provide a reasonable null model even for finite networks.
In order to test this assumption numerically, we draw networks from the directed configuration and 
compare measured quantities with basal ensemble expectations.
We note, however, that the results derived in the main text for the coherence ensemble do not 
depend on an equivalence between the basal ensemble and the directed configuration ensemble.

\setcounter{figure}{0}
\renewcommand{\figurename}{Figure S}

\begin{figure}[t!]
\begin{center}
\includegraphics[scale=0.32]{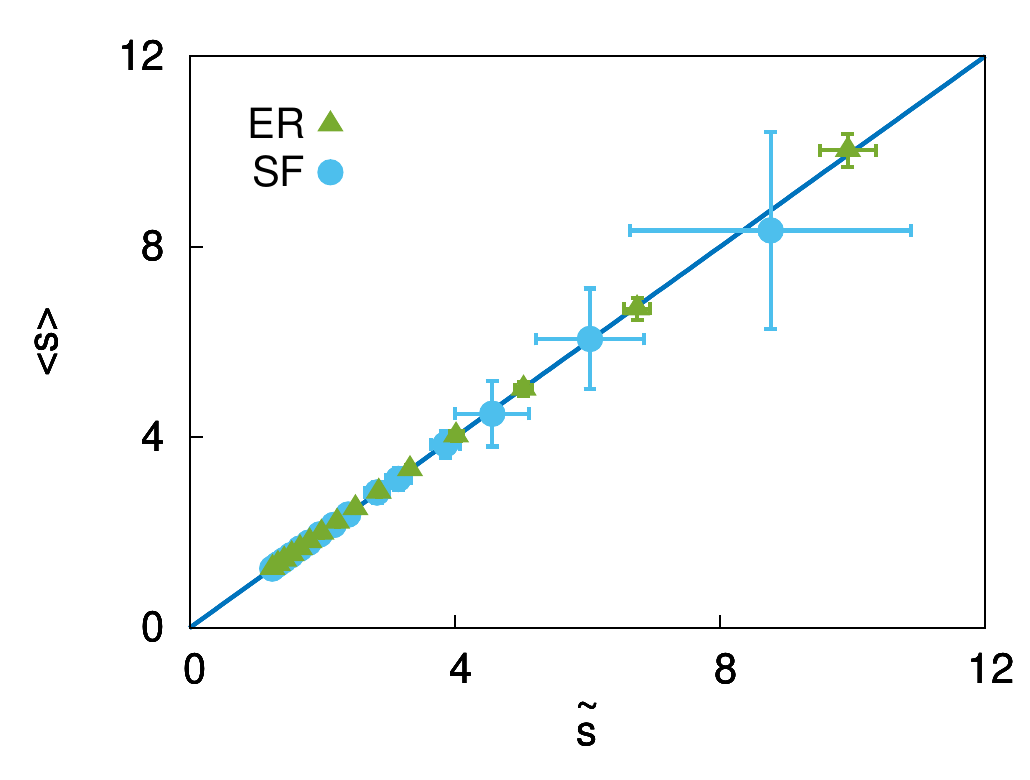}
\includegraphics[scale=0.32]{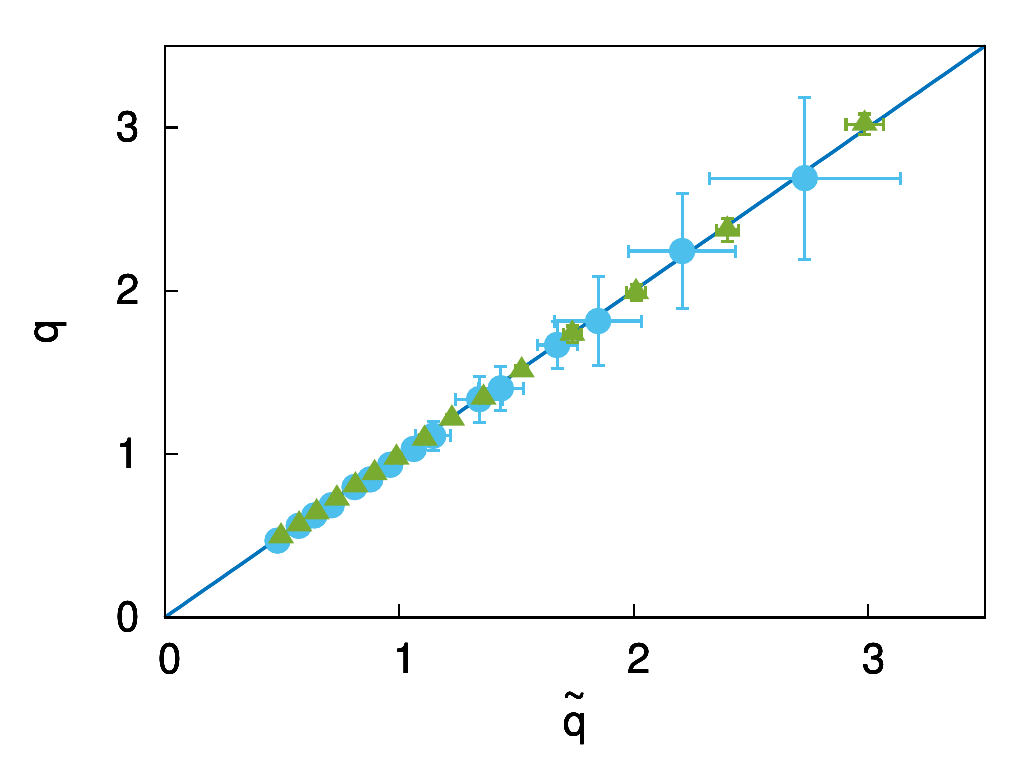}
\end{center}
\caption{
Values of mean trophic level $\langle s\rangle$ (left), and incoherence parameter $q$ (right), obtained numerically 
with two 
random graph models against the corresponding predictions for the basal ensemble, as given 
by Eqs. (\ref{eq_smed}) and (\ref{eq_qcm_mm}), respectively. The random graphs are sampled from the 
Erd\H{o}s-R\'enyi ensemble (circles) with $N=300$ and $\langle k\rangle =L/N= 10$;
and from the directed configuration ensemble (triangles) with $N=500$, $\langle k\rangle =L/N= 5$ and power-law degree sequences
of exponent $\gamma=3$. In both cases, the proportion of basal nodes, $B/N$, is varied from $10\%$ (upper right hand corners) 
to $50\%$ (lower left hand corners).
}
\label{fig_qer}
\end{figure}

%
%
%
%
%
%
%

Figure S\ref{fig_qer} displays the results of Monte Carlo simulations which support the conjecture that the basal ensemble
provides a good approximation to more general random graph ensembles as regards trophic structure.
We consider a directed version of the Erd\H{o}s-R\'enyi ensemble,
in which $L$ directed edges are distributed randomly among $N$ nodes with the constraint that there must be $B$ basal nodes.
We also take the directed configuration ensemble of networks, which restricts both the in- and out-degrees of each node 
to specified values, and generate heavy-tailed (scale-free) networks by drawing those values from independent distributions 
$p(k)\sim k^{-\gamma}$, for $k=k^{in}$ and $k=k^{out}$.
The left panel of Fig. S\ref{fig_qer} shows the mean trophic level $\langle s\rangle$ obtained numerically from Erd\H{o}s-R\'enyi networks 
and heavy-tailed networks, as described, for varying proportions of basal nodes. This is plotted against
the corresponding values given by Eq. (\ref{eq_smed}) in each case. 
Similarly, the right panel of Fig. S\ref{fig_qer} shows the incoherence parameter $q$ against the 
value given by Eq. (\ref{eq_qcm_mm}), for the same networks. In both cases, 
the results fall very close to the $f(x)=x$ line, suggesting that the trophic structure of random graphs,
regardless of their degree heterogeneity, is well approximated by the basal ensemble we have defined above.


\section{Networks with self-cycles}

\begin{figure}[ht!]
\begin{center}
\includegraphics[scale=0.6]{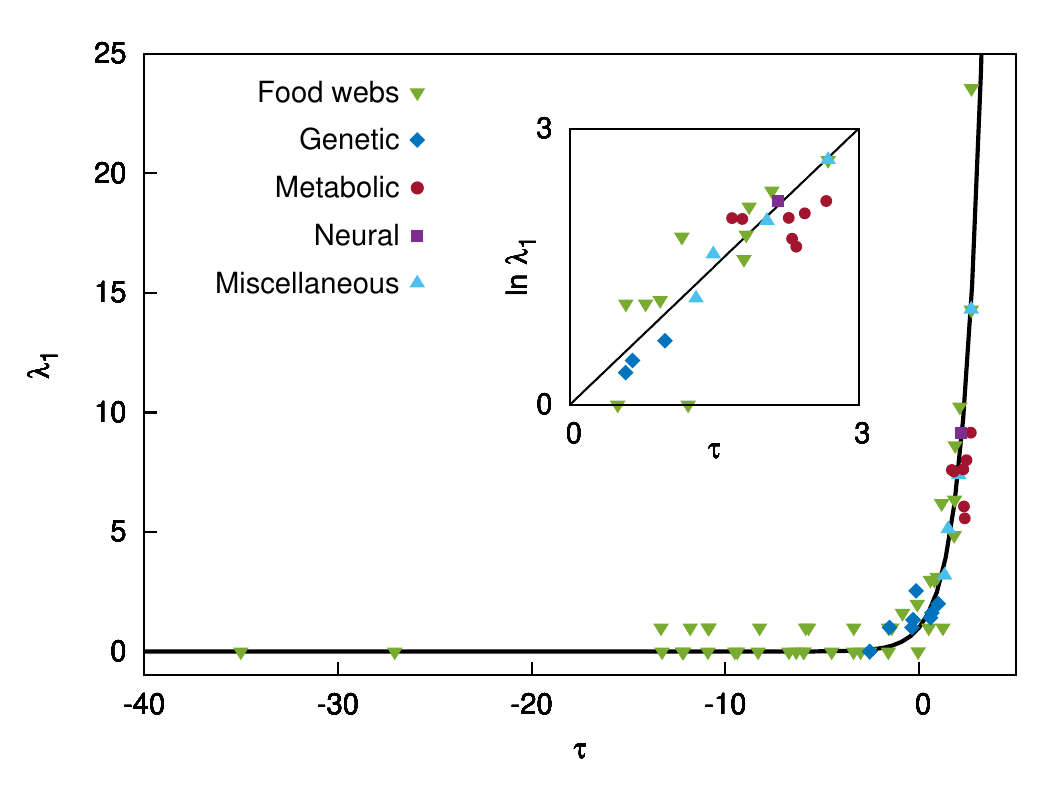}
\end{center}
\caption{
Leading eigenvalues $\lambda_1$ of several directed networks when self-edges are not excluded, 
against $\tau$ as given by Eq. (\ref{eq_tau});
symbols indicate food webs (green down-pointing triangles), gene regulatory networks (dark blue diamonds), metabolic networks 
(burgundy circles), a neural network (purple square), and other miscellaneous networks (light blue up-pointing triangles). 
Line: Expected leading eigenvalue $\overline{\lambda_1}$ in the coherence ensemble, as given by Eq. (\ref{eq_lambda_1}).
Inset: Semi-log version of the positive quadrant of the main panel (Pearson's correlation coefficient: $r^2=0.80$).
Compare with Fig. 2 of the main text, for which self-edges are excluded.
Details for each network, including references, are listed in Tables S\ref{table_foodwebs}, S\ref{table_gene_nets}.
S\ref{table_meta_nets} and S\ref{table_misc_nets} (though note that in these tables self-edges are excluded).}
\label{fig_tau}
\end{figure}

In the main text we ignore self-edges (cycles of length one) in those networks which exhibit them; this is mainly
because self-edges are not
reported in all networks, the nature of self-interaction often being considered fundamentally different 
to that of inter-element interaction. However, for completeness we also compute the values of 
the leading eigenvalue
$\lambda_1$,
and of the loop exponent
$\tau$, defined as
\begin{equation}
\tau=\ln \alpha + \frac{1}{2\tilde{q}^2}-\frac{1}{2q^2},
\label{eq_tau}
\end{equation}
when self-edges are allowed, and display these values in Fig. S\ref{fig_tau}
(compare with Fig. 1 of the main text). We can observe that the good fit to the expression
\begin{equation}
\overline{\lambda_1}=e^{\tau}
\label{eq_lambda_1}
\end{equation}
-- Eq. (14) of the main text, where $\overline{\lambda_1}$ is the coherence ensemble expectation for $\lambda_1$  --
is not significantly affected by the inclusion of self-edges. The main difference is that several of the 
food webs which have $\lambda_1=0$ when self-edges are excluded now have $\lambda_1=1$, as a result of 
cannibalism.

\section{Network data}


In the main text we assess the validity of our analytical results through comparisons with a set of empirically-derived directed networks. 
Most of these are available online, but a few of them were shared with us in private correspondence.
Below we list the most relevant details for each of these networks. Table \ref{table_foodwebs} is for 42 food webs,
Table S\ref{table_gene_nets} lists eight gene regulatory networks, and Table S\ref{table_meta_nets} contains 
information on seven metabolic networks. Table S\ref{table_misc_nets} is for six networks of various other kinds:
the neural network of {\it C. elegans},
a P2P file sharing network, two networks of trade between nations (one of basic manufactured good and the other of minerals)
and one of concatenated English words in the book {\it Green Eggs and Ham}, by Dr. Seuss;
the last of these was obtained from the original text 
for this work \cite{DrSeuss}.
The adjacency matrices of these networks are available at:\\
\url{
http://www2.warwick.ac.uk/fac/sci/maths/people/staff/sjohnson}\\
or upon request from the authors.

For each network, we list
the number of nodes $N$, the number of basal nodes $B$, and the mean degree $\langle k\rangle$;
the incoherence parameter $q$ and its ratio to the expected value $\tilde{q}$ (low ratios mean more coherent networks than randomly expected); 
the mean trophic level
$\langle s\rangle$ and the $k^{in}$-$k^{out}$ correlation parameter $\alpha$, both normalised by their
expected values; the loop exponent $\tau$,
whose sign determines whether a network is in the high or low feedback regimes (see main text); 
the leading eigenvalue, $\lambda_1$, of the adjacency matrix; and, finally, references to the sources of the data.

\begin{figure}[t!]
\begin{center}
\includegraphics[scale=0.11]{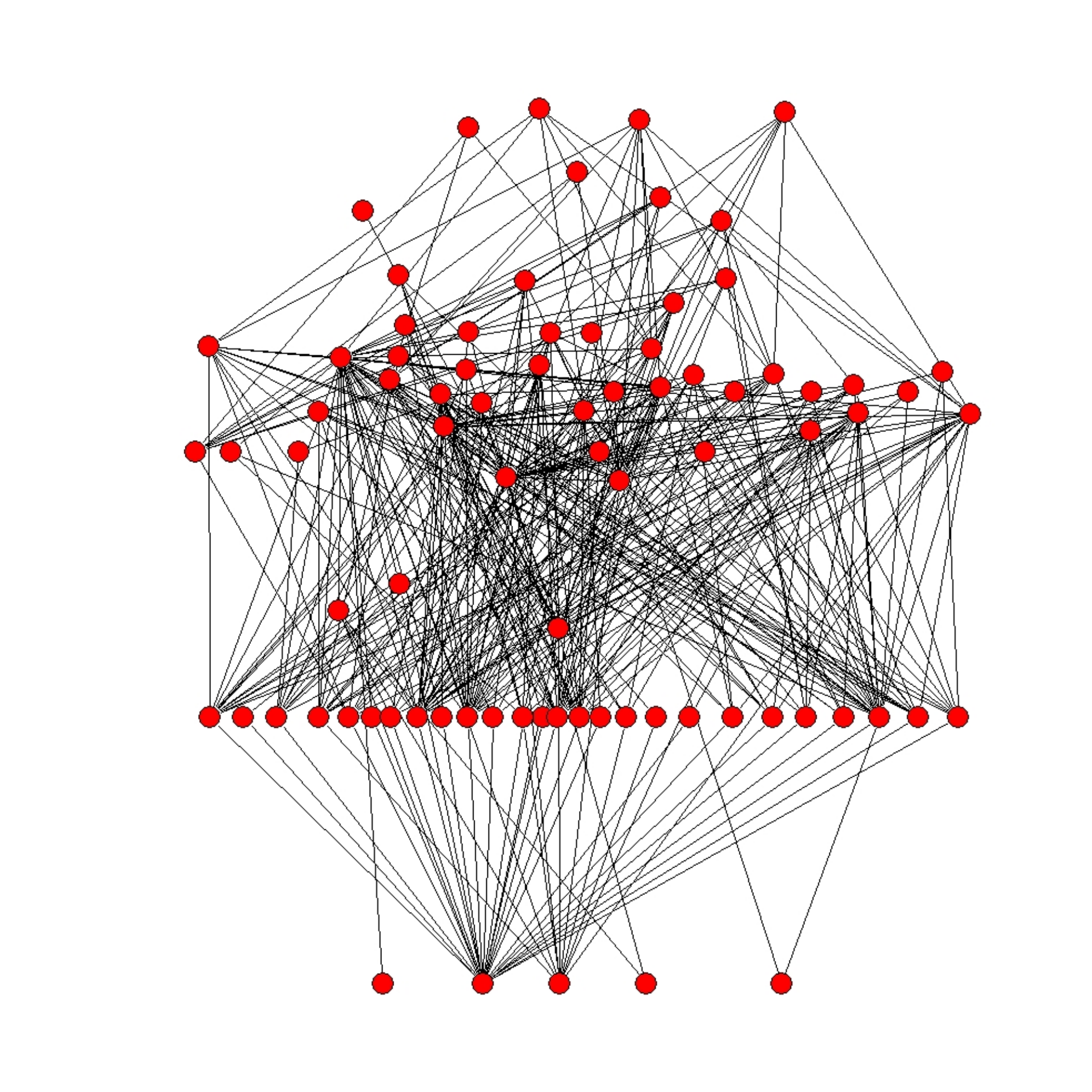}
\includegraphics[scale=0.12]{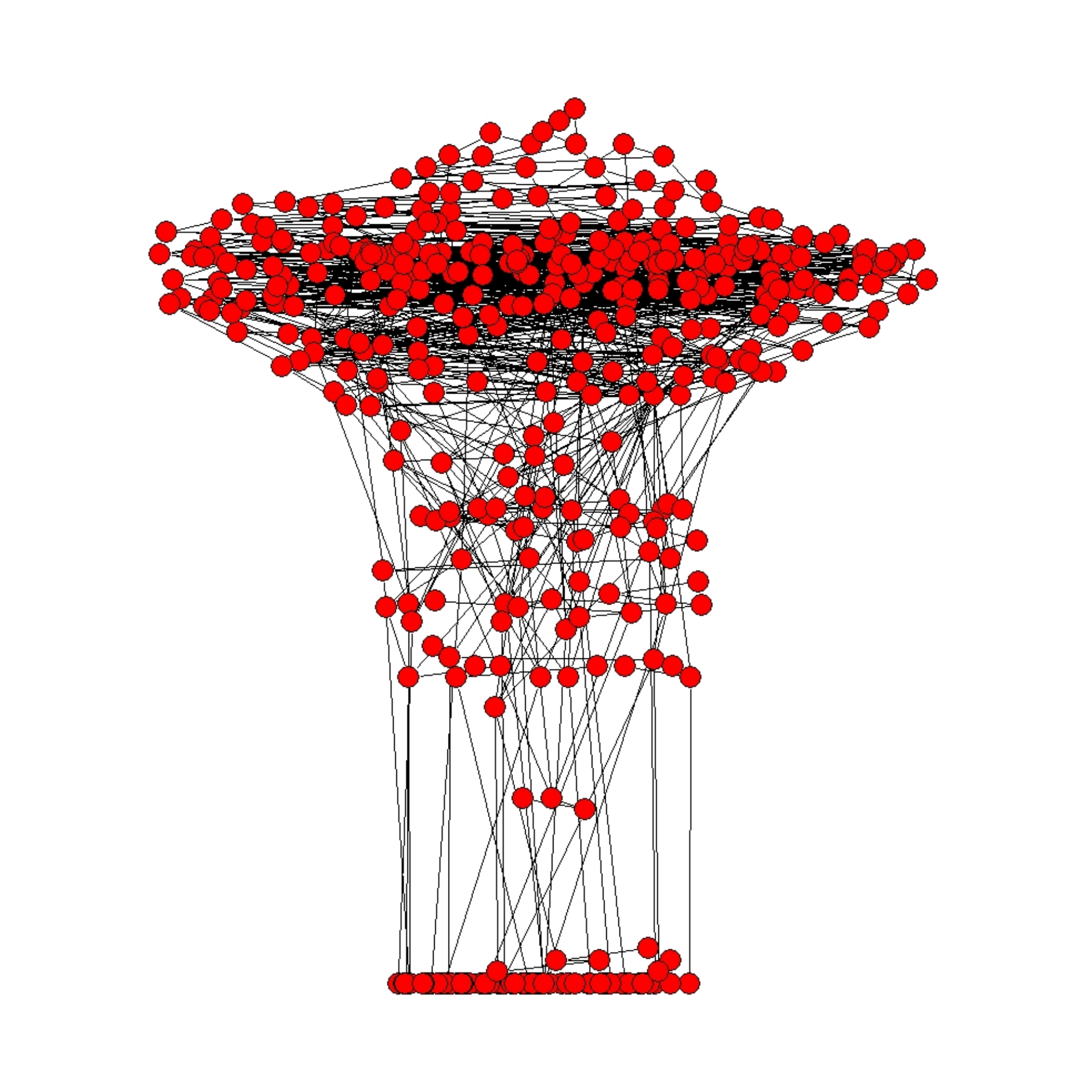}
\includegraphics[scale=0.11]{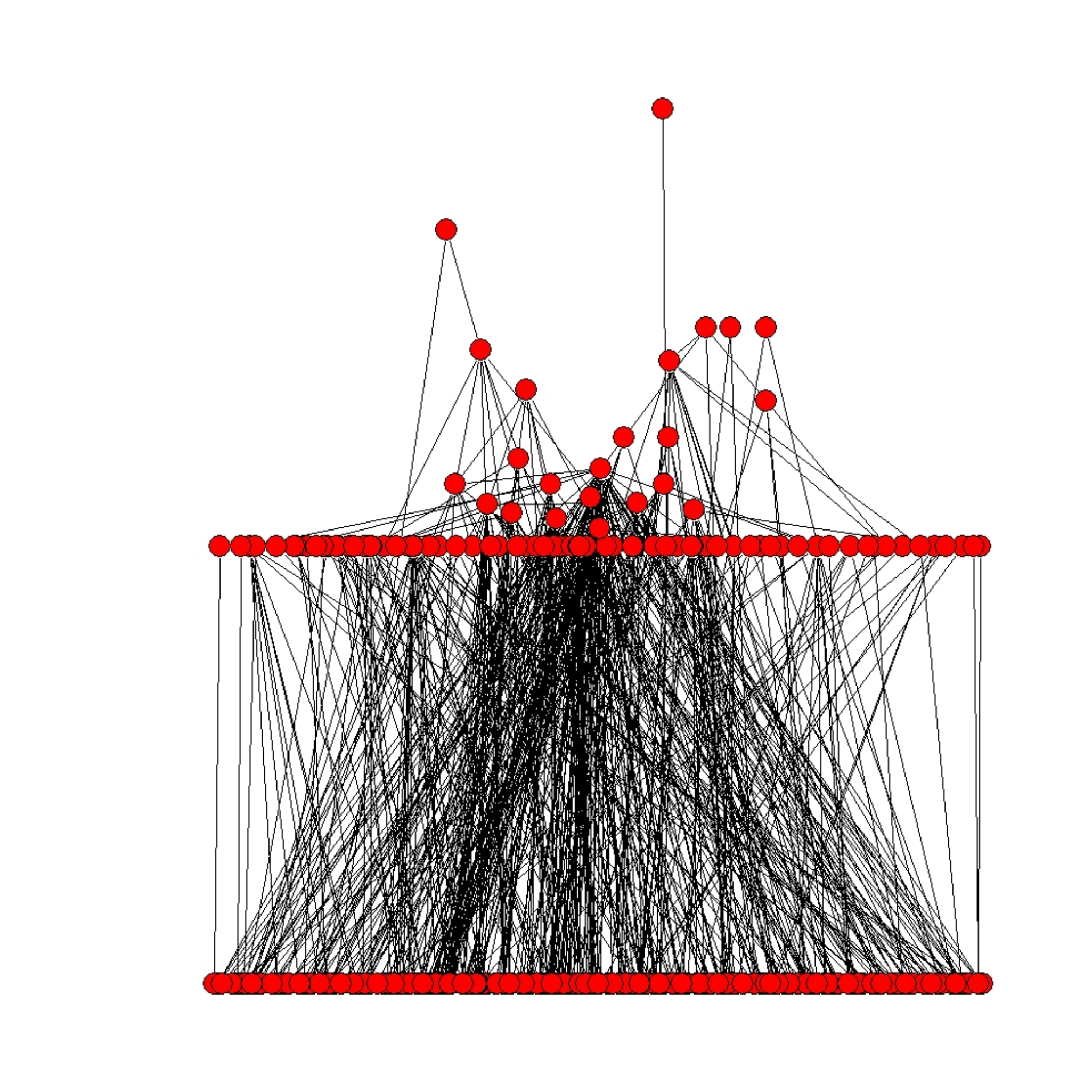}
\end{center}
\caption{
Empirical networks display a rich diversity of trophic structures.
Examples of three kinds of network are plotted here in such a way that the height of each node on the vertical axis is 
proportional to its trophic level (the scale used for each network is different because of the disparity in mean trophic level).
Left: Ythan Estuary food web \cite{Ythan96}, which is significantly more trophically coherent than the random expectation
($q/\tilde{q}=0.147$) and has no significant $k^{in}$-$k^{out}$ correlations ($\alpha/\tilde{\alpha}=0.935$);
it is in the negative $\tau$ regime: $\tau=-1.319$.
Centre: A network derived from observations of the 
{\it Chlamydia pneumoniae} metabolism \cite{jeong2000large}, which is significantly less trophically coherent
than the random expectation ($q/\tilde{q}=1.621$), and has positive $k^{in}$-$k^{out}$ correlations 
($\alpha/\tilde{\alpha}=2.550$);
it is in the positive $\tau$ regime: $\tau=1.686$.
Right: A network derived from gene regulation in {\it E. coli} \cite{ecoli_thieffry}, which is only slightly more trophically coherent than
the random expectation
($q/\tilde{q}=0.878$)
and has no significant $k^{in}$-$k^{out}$ correlations
($\alpha/\tilde{\alpha}=0.938$);
it is in the negative $\tau$ regime:
$\tau=-2.543$.
Details for each network, including references, are listed in the tables of SI.
}
\label{fig_pics}
\end{figure}

Figure S\ref{fig_pics} shows three example networks, one from each class: the Ythan Estuary food web \cite{Ythan96}, 
a metabolic network derived for 
{\it Chlamydia pneumoniae} \cite{jeong2000large}, and 
a gene regulatory network derived for {\it E. coli} \cite{ecoli_thieffry}. 
The height of each node on the vertical axis is proportional to its trophic level, and this visualisation is enough 
to show that the trophic structures of these systems can be highly informative.
Note, in particular, that a network can display significant trophic coherence
(the Ythan Estuary food web has $q/\tilde{q}=0.15$) without all its nodes falling into clearly defined trophic levels,
while it is also possible to be almost bipartite, as in the case of {\it E. coli}'s gene regulatory network,
yet be less significantly coherent as compared to the random
expectation  ($q/\tilde{q}=0.88$).
These examples show that in order to determine which regime (of high or low feedback) a given system belongs to, 
it is insufficient to look only at trophic coherence: one must compute the loop exponent $\tau$.


\renewcommand{\tablename}{Table S} 
\setcounter{table}{0}

\begin{center}
 \begin{longtable}{@{}llccccccccc}
Food web  & $N$ & $B$ & $\langle k\rangle$ & $q$ & $q/\tilde{q}$ & $\langle s\rangle/\tilde{s}$ & $\alpha/\tilde{\alpha}$ & $\tau$ & $\lambda_1$ & Ref.\\  
\hline
\hline
Benguela Current	&	29	&	2	&	6.76	&	0.69	&	0.15	&	0.17	&	0.69	&	0.50	&	2	&	\cite{benguela}	\\
Berwick Stream	&	77	&	35	&	3.12	&	0.18	&	0.53	&	1.05	&	1.06	&	-12.21	&	0	&	\cite {streams5,streams6,streams7}	\\
Blackrock Stream	&	86	&	49	&	4.36	&	0.19	&	0.57	&	1.02	&	1.27	&	-9.51	&	0	&	\cite {streams5,streams6,streams7}	\\
Bridge Brook Lake	&	25	&	8	&	5.08	&	0.53	&	0.36	&	0.68	&	0.72	&	-0.53	&	1	&	\cite{bridge}	\\
Broad Stream	&	94	&	53	&	6.00	&	0.14	&	0.49	&	1.05	&	1.20	&	-20.10	&	0	&	\cite {streams5,streams6,streams7}	\\
Canton Creek	&	102	&	54	&	6.82	&	0.15	&	0.57	&	1.01	&	1.22	&	-14.52	&	0	&	\cite{canton}	\\
Caribbean Reef	&	50	&	3	&	10.70	&	0.94	&	0.33	&	0.36	&	0.96	&	1.73	&	7.80	&	\cite{reef}	\\
Cayman Islands	&	242	&	10	&	15.55	&	0.77	&	0.24	&	0.30	&	0.51	&	1.22	&	0	&	\cite{caribbean_2005}	\\
Catlins Stream	&	48	&	14	&	2.29	&	0.20	&	0.41	&	0.98	&	1.00	&	-10.90	&	0	&	\cite {streams5,streams6,streams7}	\\
Chesapeake Bay	&	31	&	5	&	2.16	&	0.45	&	0.33	&	0.73	&	0.90	&	-1.81	&	0	&	\cite{chesapeake1,chesapeake2}	\\
Coachella Valley	&	29	&	3	&	8.38	&	1.20	&	0.48	&	0.47	&	0.91	&	1.63	&	5.48	&	\cite{coachella}	\\
Coweeta 1	&	58	&	28	&	2.17	&	0.30	&	0.64	&	1.00	&	1.08	&	-3.39	&	0	&	\cite {streams5,streams6,streams7}	\\
Coweeta 17	&	71	&	38	&	2.08	&	0.24	&	0.60	&	1.00	&	1.25	&	-5.94	&	0	&	\cite {streams5,streams6,streams7}	\\
Dempsters (Au)	&	83	&	46	&	4.99	&	0.21	&	0.57	&	1.08	&	1.02	&	-7.42	&	0	&	\cite {streams5,streams6,streams7}	\\
Dempsters (Sp)	&	93	&	50	&	5.78	&	0.13	&	0.38	&	1.07	&	1.11	&	-27.07	&	0	&	\cite {streams5,streams6,streams7}	\\
Dempsters (Su)	&	107	&	50	&	9.02	&	0.27	&	0.57	&	1.05	&	1.04	&	-3.51	&	0.01	&	\cite {streams5,streams6,streams7}	\\
El Verde Rainforest	&	155	&	28	&	9.72	&	1.01	&	0.45	&	0.49	&	1.21	&	2.09	&	10.12	&	\cite{el_verde}	\\
German Stream	&	84	&	48	&	4.19	&	0.20	&	0.47	&	1.02	&	1.10	&	-9.35	&	0	&	\cite {streams5,streams6,streams7}	\\
Healy Stream	&	96	&	47	&	6.60	&	0.22	&	0.53	&	1.03	&	1.12	&	-6.34	&	0	&	\cite {streams5,streams6,streams7}	\\
Kyeburn Stream	&	98	&	58	&	6.42	&	0.18	&	0.62	&	1.02	&	1.18	&	-9.39	&	0	&	\cite {streams5,streams6,streams7}	\\
LilKyeburn Stream	&	78	&	42	&	4.81	&	0.23	&	0.53	&	1.01	&	1.10	&	-5.97	&	0	&	\cite {streams5,streams6,streams7}	\\
Little Rock Lake	&	92	&	12	&	10.7	&	0.67	&	0.22	&	0.25	&	0.77	&	1.06	&	5.66	& 	\cite{little_rock}	\\
Lough Hyne	&	349	&	49	&	14.62	&	0.60	&	0.37	&	0.59	&	0.63	&	0.85	&	2.56	&	\cite{lough_hyne_1,lough_hyne_2}	\\
Martins Stream	&	105	&	48	&	3.27	&	0.32	&	0.58	&	0.99	&	1.26	&	-2.56	&	0	&	\cite {streams5,streams6,streams7}	\\
Narrowdale Stream	&	71	&	28	&	2.17	&	0.23	&	0.50	&	0.98	&	1.17	&	-7.45	&	0	&	\cite {streams5,streams6,streams7}	\\
NE Shelf	&	79	&	2	&	17.44	&	0.73	&	0.13	&	0.10	&	0.71	&	1.57	&	4.32	&	\cite{shelf}	\\
North Col Stream	&	78	&	25	&	3.09	&	0.28	&	0.52	&	0.98	&	1.36	&	-4.52	&	0	&	\cite {streams5,streams6,streams7}	\\
Powder Stream	&	78	&	32	&	3.44	&	0.22	&	0.47	&	0.99	&	1.12	&	-8.32	&	0	&	\cite {streams5,streams6,streams7}	\\
Scotch Broom	&	85	&	1	&	2.58	&	0.40	&	0.14	&	0.30	&	1.20	&	-2.08	&	0	&	\cite{broom}	\\
Skipwith Pond	&	25	&	1	&	7.56	&	0.61	&	0.15	&	0.16	&	0.64	&	0.20	&	2	&	\cite{skipwith}	\\
St Marks Estuary	&	48	&	6	&	4.54	&	0.63	&	0.37	&	0.63	&	1.02	&	0.26	&	0	&	\cite{st_marks}	\\
St Martin Island	&	42	&	6	&	4.88	&	0.59	&	0.32	&	0.54	&	0.79	&	-0.05	&	0.01	&	\cite{st_martin}	\\
Stony Stream	&	109	&	61	&	7.59	&	0.15	&	0.55	&	1.03	&	1.16	&	-14.66	&	0	&	\cite{stony}	\\
Stony Stream 2	&	112	&	63	&	7.41	&	0.15	&	0.55	&	1.04	&	1.18	&	-14.72	&	0	&	\cite {streams5,streams6,streams7}	\\
Sutton (Au)	&	80	&	49	&	4.19	&	0.15	&	0.66	&	1.08	&	1.28	&	-13.27	&	0	&	\cite {streams5,streams6,streams7}	\\
Sutton (Sp)	&	74	&	50	&	5.28	&	0.10	&	0.56	&	1.11	&	1.15	&	-35.01	&	0	&	\cite {streams5,streams6,streams7}	\\
Sutton (Su)	&	87	&	63	&	4.87	&	0.28	&	0.89	&	1.19	&	0.52	&	-1.59	&	0	&	\cite {streams5,streams6,streams7}	\\
Troy Stream	&	77	&	40	&	2.35	&	0.19	&	0.37	&	1.01	&	1.14	&	-12.16	&	0	&	\cite {streams5,streams6,streams7}	\\
UK Grassland	&	61	&	8	&	1.59	&	0.40	&	0.18	&	0.42	&	0.63	&	-3.03	&	0	&	\cite{grass}	\\
Venlaw Stream	&	66	&	30	&	2.83	&	0.23	&	0.54	&	1.06	&	1.35	&	-6.72	&	0	&	\cite {streams5,streams6,streams7}	\\
Weddel Sea	&	483	&	61	&	31.71	&	0.72	&	0.55	&	0.75	&	1.17	&	2.63	&	22.91	&	\cite{weddell_sea}	\\
Ythan Estuary	&	82	&	5	&	4.77	&	0.42	&	0.15	&	0.28	&	0.93	&	-1.32	&	1	&	\cite{Ythan96}	\\
\caption{
 Details of 42 food webs used in the main text.
 Columns are for number of nodes $N$, number of basal nodes $B$,
 mean degree $\langle k\rangle$, and
incoherence parameter $q$;
ratios of $q$, mean trophic level $\langle s\rangle$, and correlation parameter $\alpha$ to their expected values in the
basal ensemble;
loop exponent $\tau$, leading eigenvalue $\lambda_1$, and references to the data sources.
Many of the data are available online at:
\url{https://www.nceas.ucsb.edu/interactionweb/html/thomps_towns.html} 
}
\label{table_foodwebs}
 \end{longtable}
\end{center}

\begin{center}
 \begin{longtable}{@{}llccccccccc}
Gene regulatory network  & $N$ & $B$ & $\langle k\rangle$ & $q$ & $q/\tilde{q}$ & $\langle s\rangle/\tilde{s}$ & $\alpha/\tilde{\alpha}$ & $\tau$ & $\lambda_1$ & Ref.\\    
\hline
\hline
Human (healthy)	&	4071	&	4004	&	2.08	&	0.08	&	0.99	&	1.00	&	0.99	&	-1.54	&	1	&	\cite{human_gerstein,BQS}	\\
Human (cancer)	&	4049	&	3967	&	2.89	&	0.08	&	1.00	&	1.00	&	1.07	&	-0.16	&	2.54	&	\cite{human_gerstein,BQS}	\\
{\it E. coli} (Salgado)	&	1470	&	1316	&	1.98	&	0.23	&	1.03	&	1.00	&	1.21	&	0.65	&	1.62	&	\cite{ecoli_salgado,BQS}	\\
{\it E. coli} (Thieffry)	&	418	&	312	&	1.24	&	0.27	&	0.88	&	1.01	&	0.94	&	-2.54	&	0	&	\cite{ecoli_thieffry,UriAlon}	\\
{\it S. cerevisiae} (Harbison)	&	2933	&	2764	&	2.10	&	0.17	&	0.98	&	1.00	&	1.29	&	-0.38	&	1	&	\cite{yeast_harbison,BQS}	\\
{\it S. cerevisiae} (Costanzo)	&	688	&	557	&	1.57	&	0.25	&	1.04	&	1.00	&	0.81	&	-0.31	&	1.32	&	\cite{yeast_costanzo,UriAlon}	\\
{\it P. aeruginosa}	&	691	&	606	&	1.43	&	0.30	&	1.00	&	1.03	&	1.94	&	0.58	&	1.41	&	\cite{paeruginosa_galan,BQS}	\\
{\it M. tuberculosis}	&	1624	&	1542	&	1.95	&	0.17	&	1.02	&	1.00	&	1.24	&	0.99	&	2.00	&	\cite{mtuberculosis_sanz,BQS}	\\
\caption{
Details of eight gene regulatory networks (GRN) used in the main text.
The E. coli (Salgado) and Yeast (Harbison) are available online at: \url{http://wws.weizmann.ac.il/mcb/UriAlon/download/collection-complex-networks}.
The others were shared with us by Luca Albergante, and some of them can be obtained from various websites: \url{http://regulondb.ccg.unam.mx/} (E. coli, Salgado);
\url{http://younglab.wi.mit.edu/regulatory_code} (Yeast, Harbison);
\url{http://www.genome.gov/ENCODE/} (Human, both the non-cancer GM12878 cell line and the K562 leukaemia cell line).
Columns as in Table S\ref{table_foodwebs}.
}
\label{table_gene_nets}
 \end{longtable}
\end{center}

\newpage

\begin{center}
 \begin{longtable}{@{}llccccccccc}
Metabolic network  & $N$ & $B$ & $\langle k\rangle$ & $q$ & $q/\tilde{q}$ & $\langle s\rangle/\tilde{s}$ & $\alpha/\tilde{\alpha}$ & $\tau$ & $\lambda_1$ & Ref.\\  
\hline
\hline
{\it A. fulgidus}	&	1267	&	36	&	2.38	&	13.79	&	1.88	&	2.06	&	4.34	&	2.35	&	7.62	&	\cite{jeong2000large}	\\
{\it M. thermoautotrophicum}	&	1111	&	30	&	2.43	&	12.17	&	1.77	&	1.90	&	4.08	&	2.31	&	7.59	&	\cite{jeong2000large}	\\
{\it M. jannaschii}	&	1081	&	32	&	2.40	&	12.47	&	1.86	&	1.98	&	4.00	&	2.27	&	7.53	&	\cite{jeong2000large}	\\
{\it C. pneumoniae}	&	386	&	20	&	2.05	&	8.98	&	1.62	&	1.71	&	2.55	&	1.69	&	5.57	&	\cite{jeong2000large}	\\
{\it C. trachomatis}	&	446	&	19	&	2.11	&	11.77	&	1.95	&	2.02	&	2.77	&	1.79	&	6.07	&	\cite{jeong2000large}	\\
{\it S. cerevisiae} (yeast)	&	1510	&	43	&	2.54	&	14.61	&	1.73	&	1.82	&	5.54	&	2.66	&	9.15	&	\cite{jeong2000large}	\\
{\it C. elegans}	&	1172	&	40	&	2.44	&	13.29	&	1.86	&	2.04	&	4.60	&	2.44	&	8.00	&	\cite{jeong2000large}	\\
\caption{
Details of seven metabolic networks used in the main text, downloaded from 
\url{http://www3.nd.edu/~networks/resources.htm}.
Columns as in Table S\ref{table_foodwebs}.
}
\label{table_meta_nets}
 \end{longtable}
\end{center}

\begin{center}
 \begin{longtable}{@{}llccccccccc}
Network (miscellaneous)  & $N$ & $B$ & $\langle k\rangle$ & $q$ & $q/\tilde{q}$ & $\langle s\rangle/\tilde{s}$ & $\alpha/\tilde{\alpha}$ & $\tau$ & $\lambda_1$ & Ref.\\ 
\hline
\hline
Neural ({\it C. elegans})	&	297	&	3	&	7.90	&	1.49	&	0.42	&	0.39	&	1.42	&	2.17	&	9.15	&	\cite{CElegans_neural,WattsStrogatz}	\\
P2P (Gnutella 2008)	&	6301	&	3836	&	3.30	&	0.98	&	0.98	&	1.00	&	1.07	&	1.49	&	5.12	&	\cite{Gnutella_1,Gnutella_2}	\\
Trade (manufactured goods)	&	24	&	2	&	12.92	&	4.24	&	1.14	&	1.14	&	1.10	&	2.68	&	14.3	&	\cite{Pajek_trade}	\\
Trade (minerals)	&	24	&	3	&	5.63	&	4.04	&	1.02	&	0.97	&	1.28	&	2.05	&	7.38	&	\cite{Pajek_trade}	\\
Words	&	50	&	16	&	2.02	&	2.04	&	1.01	&	1.16	&	1.55	&	1.31	&	3.17	&	\cite{DrSeuss}	\\
\caption{
Details of six other networks used in the main text.
The network of words was obtained for this work from
{\it Green Eggs and Ham} \cite{DrSeuss}, and is available upon request from s.johnson.2@warwick.ac.uk.
The other data can be found on various websites:
\url{http://www-personal.umich.edu/~mejn/netdata/} (neural network);
\url{https://snap.stanford.edu/data/p2p-Gnutella08.html} (P2P network); and
\url{http://vlado.fmf.uni-lj.si/pub/networks/data/esna/metalWT.htm} (trade networks).
Columns as in Table S\ref{table_foodwebs}
}
\label{table_misc_nets}
 \end{longtable}
\end{center}






\subsection{Green Eggs and Ham words network}

\begin{figure}[ht!]
\begin{center}
\includegraphics[scale=0.28]{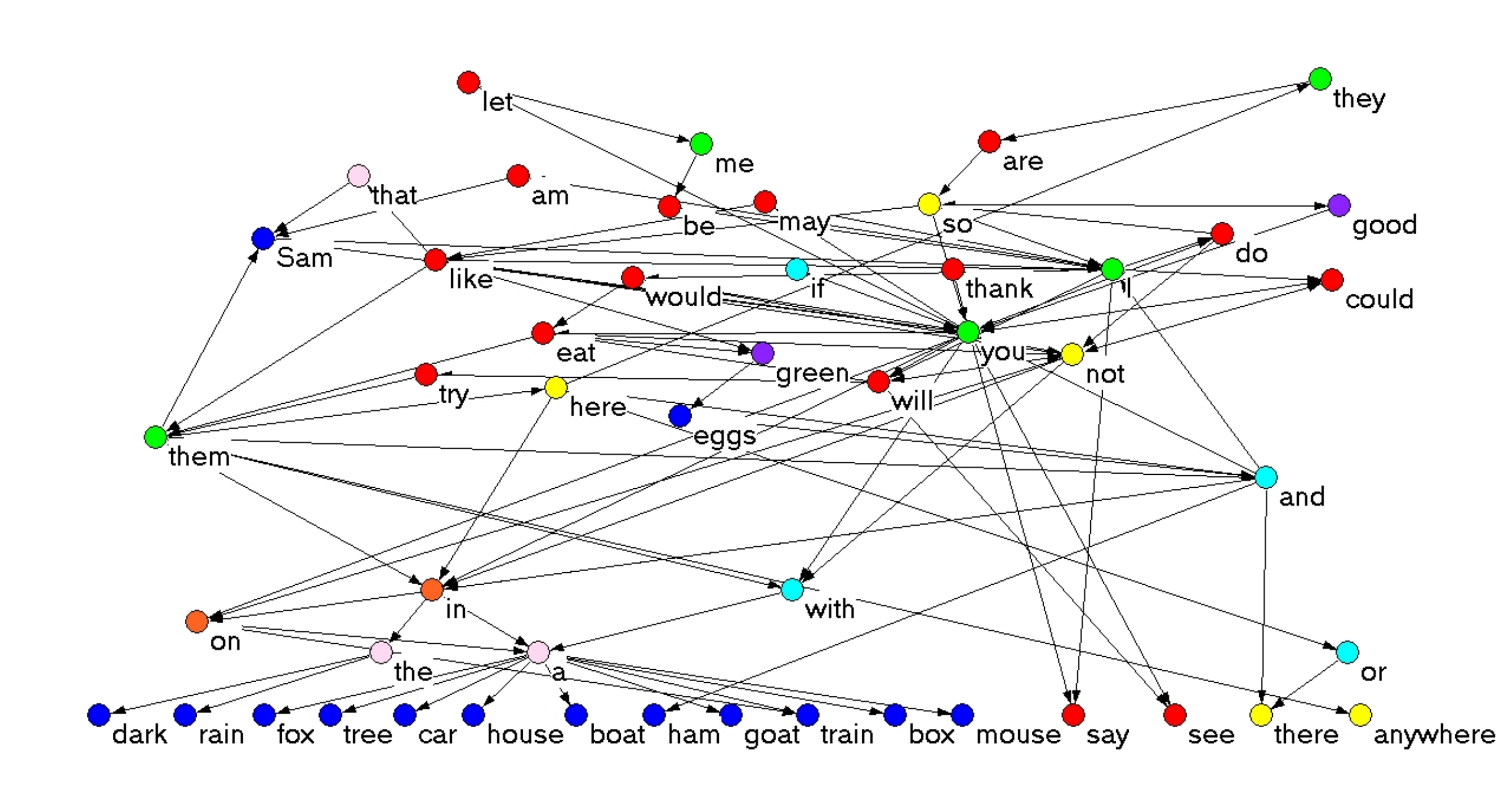}
\end{center}
\caption{\small{Network of concatenated words from {\it Green Eggs and Ham}, by Dr Seuss \cite{DrSeuss}.
The height of each word is proportional to its trophic level.
Colours indicate syntactic function;
from lowest to highest mean trophic level: nouns (blue), prepositions and conjunctions (cyan), determiners (pink),
adverbs (yellow), pronouns (green), verbs (red), and adjectives (purple). 
When a word has more than one function, the one most common in the text is used.
}}
\label{fig_green_eggs}
\end{figure}

We obtained the network of concatenated words from Dr Seuss's masterpiece {\it Green Eggs and Ham} in the following way \cite{DrSeuss}.
Every node in the text was assigned a node, and a directed edge $a_{ij}=1$ was placed whenever word $i$ preceded word $j$ in a sentence.
Figure S\ref{fig_green_eggs} displays this network in such a way that the height of each node on the y-axis is proportional to its trophic level.
Note that the arrows we have placed between nodes in this visualisation are from the preceding word to the succeeding one, so that one can 
obtain sentences (some of them grammatically correct) by following the arrows.
We have used colours to indicate syntactic function, as described in the caption.

\newpage
